%% file: document.tex
\documentclass{bioinfo}
\copyrightyear{2014}
\pubyear{2014}
\usepackage{amsfonts}
\usepackage{booktabs}
\usepackage{csvsimple}

\usepackage{mathtools}    % loads »amsmath«
\DeclareMathOperator*{\argmin}{argmin}

\begin{document}
\firstpage{1}

%\title[short Title]{Are 20 years of LASSO sufficient to infer gene interaction networks?}
%\title[LASSO, 20 years later]{LASSO, 20 years later. How far are we from correctly inferring gene interaction networks?}
\title[Lasso for gene interaction networks]{Are we far from correctly inferring gene interaction networks with Lasso?}
\author[Gadaleta \textit{et~al}]{Francesco Gadaleta\,$^{1}$\footnote{to whom correspondence should be addressed} }
\address{$^{1}$ University of Leuven, Belgium }

\history{
%Received on XXXXX; revised on XXXXX; accepted on XXXXX
}
\editor{
%Associate Editor: XXXXXXX
}
\maketitle

\input{abstract}
\input{introduction}
\input{methods}
\input{comparison}
\input{discussion}
\input{conclusion}

%\section*{Acknowledgement}
\paragraph{Funding\textcolon} 

\bibliographystyle{abbrv}
\bibliography{Document}
\end{document}

%% file: abstract.tex
\begin{abstract}
Detecting the interactions of genetic compounds like genes, SNPs, proteins, metabolites, etc. can potentially unravel the mechanisms behind complex traits and common genetic disorders. Several methods have been taken into consideration for the analysis of different types of genetic data, regression being one of the most widely adopted. Without any doubt, a common data type is represented by gene expression profiles, from which gene regulatory networks have been inferred with different approaches. In this work we review nine penalised regression methods applied to microarray data to infer the topology of the network of interactions. We evaluate each method with respect to the complexity of biological data. We analyse the limitations of each of them in order to suggest a number of precautions that should be considered to make their predictions more significant and reliable.

\section{Contact:} \href{francesco.gadaleta@gmail.com}{francesco.gadaleta@gmail.com}
\end{abstract}

%% file: introduction.tex
\section{Introduction}
Complexity of biological systems is dictated by their interactions. In the field of gene regulatory networks, detecting significant interactions means understanding the biological mechanisms that regulate complex genetic disorders.
%by detecting the genetic compounds that are likely to be related to the disease of interest. 
Graphical models are a common mathematical abstraction that allow researchers to visualise those interactions, detect groups of similar predictor variables, discover pathways or assess the conditional dependency between covariates. All this information is at the researcher's disposal by means of graphs, formed by nodes, edges and the connections between them via the adjacency matrix. The values of each entry $(i, j)$ in such a matrix indicate the magnitude of the interaction between nodes $i$ and $j$. There have been several attempts to analyse biological data with graph theory. The current trends consist in regressing the phenotype of a number of individuals against their genetic profile or regressing clinical data to perform survival analysis. The types of data might be heterogeneous and can include expression profiles, RNAseq data, SNPs, proteins or metabolites.
In this work we refer to the study of genetic interaction networks and review some methods that have been specifically designed to infer the topology of such networks. We refer to inferring network topologies with methods based on penalised regression from gene expression data.

In Section 1 we introduce $L1-$norm penalised regression. In Section \ref{methods} we provide a description of methods derived from it. In Section \ref{comparison} we discuss the performance of the aforementioned methodologies and highlight the ones that perform the best on synthetic datasets.
Section \ref{conclusion} draws our conclusion, paving the way to potential improvements in terms of accuracy and computational burden whenever dealing with real-life data.

%% file: methods.tex
\section{Methods} \label{methods}
Penalised regression has been considered within several domains of computational biology. Some of these contributions include the analysis of kidney cancer microarray data, regressed against survival times of each individual \cite{citeulike:3366723}. In such a context, a Lasso method has been applied to preconditioned response variables \cite{pauL2008}. Penalised regression has been applied also in genome-wide association studies under the name of Lasso logistic regression \cite{Wu_langek}, in which main effects are analysed together with interaction effects for SNPs data, and hyper-lasso \cite{hyperlasso}. 
%%%FIXME paper is too long already
%%%Improved Feature Selection by Incorporating Gene Similarity into the Lasso
%%%\cite{Gillies2012}
A contribution within the field of neuroscience applies Lasso regression to evaluate genetic effects with respect to brain images, using MRI-derived temporal lobe volume measure as response variable \cite{neurolasso}. %Discovery and replication of gene influences on brain structure using Lasso regression (not gene interactions just application of lasso)
%1 
Several attempts to improve the performance of the Lasso procedure led to a very efficient algorithm developed by Friedman et al. \cite{glasso}, called graphical lasso that maximises the penalised log-likelihood function through coordinate-descent.
%2 this already added
%Genome-wide association analysis has been made possible with Lasso penalised logistic regression in \cite{Wu_langek}, in which main effects are analysed together with interaction effects for SNPs data. This approach has been implemented in the Mendel software. 
%3
%Gene-gene interactions with lasso claimed as successful (Check out these papers)
An attempt to detect gene-gene interactions with a combination of Lasso and Principal Component Analysis is provided in \cite{angelo}. 
%which combines Lasso with principal-components analysis to analyse gene expression profiles in genome-wide association studies. BMC Proc. 3(Suppl.), S7?S62. doi:10.1186/1753-6561-3- S7-S62
%4
%Additionally, the work of FIXME Li, J., Das, K., Fu, G., Li, R., and Wu, R. (2011) performs Bayesian lasso for genome-wide association studies. Bioinformatics 27, 516?523.
%Inference of Genetic Networks from Time Course Expression Data Using Functional Regression with Lasso Penalty (hong and lian functional lasso)
%5
%PENALIZED LINEAR UNBIASED SELECTION also called "Nearly unbiased variable selection under minimax concave penalty" \cite{zhang2010} (some interesting features about convexity and computational complexity)
In this specific work we focus on the application of penalised regression methods applied to variable selection and structure inference, with the purpose of discovering the network topology that regulates the interaction of genetic compounds. Within this context, seminal work of Meinshausen et al. \cite{Meinshausen06highdimensional} paved the way to a simple yet effective approach that performs Lasso iteratively on each response variable. According to their methodology, given the expression profiles of individuals known to be affected by complex genetic trait, each gene is first considered as response and regressed against the remaining ones. %Every gene is treated, in turn, as response and the regression continues until all genes have been analysed. 
The problem of iterative regression is translated into the more intuitive one of neighbourhood selection: only the genes that are directly associated to the response are selected and their coefficients estimated.  These associations can be visualised as a graph in which nodes and edges represent genes and interactions, respectively. Nevertheless, a number of limitations affect such a procedure, especially when it is applied to real life data. 
We will address a number of such limitations in Section \ref{limitation}. 

\subsection{$L1$-norm: the common form of penalised regression}
One common form of penalised regression in computational biology is the Lasso procedure introduced by Tibshirani \cite{Tibshirani94regressionshrinkage}. This procedure has an attractive feature referred to as \emph{regularisation by $L1$-norm}. 
Given a $n \times p$ $n$-dimensional matrix $X$ and a $n$-dimensional response vector $Y$, the Lasso estimate is given by

\begin{equation}
\label{eq:lasso}
    \hat{\Theta}^{a,\lambda} = 
    \argmin_{%
      \substack{%
        \text{s.\,t.}\, \Theta:\Theta_a = 0 \\
        \phantom{}\, 
      }
    }
    (\frac{1}{n} \| Y_i - X\Theta \|^{2}_2 + \lambda \| \Theta \|_1)
  \end{equation}
   
where $|| \Theta ||_1 = \sum_{j} \theta_j$ is the $L1$-norm of the coefficient vector. 

The minimisation problem of Equation \ref{eq:lasso} finds the best $Theta$ that minimises the least squares with a penalty factor. A property of the $L1$-norm of the minimisation problem above is that it tends to shrink the coefficients of a number of variables to zero. By doing so, it discards them from the set of selected variables associated to the response, making the model simpler. As a matter of fact, a simpler model is affected by smaller variance of the regressed coefficients, at the cost of an increased bias of the predicted response. However, in a variable selection procedure, a lower number of variables is usually preferred to a less biased prediction. 
In the context of microarray analysis, the terms $X$ and $Y$ of Equation \ref{eq:lasso} correspond to the $p$ gene expression profiles of $n$ individuals and the expressions of the response gene of each individual, respectively. Correctly estimating the shrinkage factor $\lambda$ is critical and challenging. In fact, the value of $\lambda$ directly determines the rate of false positives and false negatives of the predictive model. A small penalty factor will allow many more genes to be added to the model. In contrast, a larger $\lambda$ will shrink a higher number of $\theta_j$ to zero, resulting in the selection of fewer genes as influential.

One of the main reasons for which Lasso is widely accepted in the field of computational biology is because the shrinkage factor $\lambda$ has an intuitive counterpart in biological terms. The shrinkage factor fits relatively well with the widely accepted biological assumption that a small number of main effects are associated to the response \cite{Michalak2008243, YiST07}. Moreover, fewer genes make the model easier to interpret, with respect to a model with a high number of degrees of freedom. 
Unfortunately, as we will explain later in this section, Lasso does not provide consistent predictions within a large number of scenarios. To begin with, regardless the convexity of the set of solutions provided by Lasso, those are not necessarily unique. As we will see, this is an issue more and more often neglected when analysing real biological data with Lasso.

\subsection{Limitations of Lasso}\label{limitation}
In this section we explore some of the limitations of penalised regression applied to high dimensional data and more specifically to genetic data, such as gene expression profiles, gene methylation data, SNPs, CNVs, etc.
\paragraph{Uniqueness of the solution}
One problem that arises when the number of predictors exceeds the number of observations, usually referred to as $p >> n$ problems, is that the Lasso criterion is not strictly convex \cite{lassounique}. This fact leads to a fundamental consequence that is not always taken into account during the analysis: the optimisation problem might not have a unique minimum. 

Specifically, the $L1$-norm Lasso solution is unique only when $rank(X) = p$. If the $rank(X) < p$ there can be more than one minimiser of the optimisation problem of Equation \ref{eq:lasso}. This occurs whenever there is sort of a structure within the data (and $p = n$) or whenever the number of observations is higher than the number of predictors ($p>n$). 
Multiple solutions $\Theta$ that give the same fitted value $\hat{y} = X\Theta$ make it impossible to interpret the results of a Lasso regression. 
For the sake of completeness, what two different Lasso solutions cannot do is to attach opposite signs to the coefficients of the same variable. An important finding reported in \cite{lassounique} consists in the fact that a unique solution exists with probability one only if the predictors are drawn from a continuous probability distribution. Moreover, the uniqueness of solution occurs regardless of the sizes of $n$ and $p$ and the maximum number of selected predictors (the nonzero components) is $min(n,p)$. It comes without saying that such a condition is rarely fulfilled in genetics, where data might contain discrete variables or it might have been post-processed before regression. 

\paragraph{Significance}
All procedures of the Lasso family lack of the usual constructs to assess significance of estimated predictors, such as p-values or confidence intervals. One common approach to evaluate the significance of predictors relies on resampling and data splitting. The major limitation addressed by such methods is the high computational burden, which becomes prohibitive for numbers of predictors that exceed $10^4$.
The lack of a statistical significance procedure for Lasso has been partially solved by a number of methods such as the one described in \cite{wasserman2009} that estimates p-values in high dimensional models based on data splitting; two more methods that derive confidence intervals for the regression coefficients are described in \cite{potscher2010} and \cite{zhang}; a method called \emph{stability selection} that controls false positives by resampling in the space of predictors has been proposed in \cite{Meinshausen_stabilityselection}; another method that constructs p-values of predictors starting from a ridge estimate and then corrects the prediction bias with Lasso has been proposed by \cite{buhlmann2013}; and two methods that give a simple statistic of Lasso coefficients without relying on sampling nor splitting data, as described in \cite{lassostatistic} and \cite{loc13sig}.

\paragraph{Multicollinearity} occurs whenever gene expression profiles are affected by the presence of highly correlated predictors \cite{multicollinearity_kvs, farrar1964multicollinearity}.  
Multicollinearity can degrade the performance and the stability of regression estimates, giving rise to non-sensical results or incorrect magnitude and sign of regression coefficients. When the number of predictors increases, such critical scenarios gain greater chances to occur. It is widely accepted that strong genetic correlations are frequent in genetics (specifically in microarray datasets). Moreover, complete independence between gene expression measurements is rare \cite{genesets}. The assumption of functionally related genes being correlated to each other is realistic. Therefore, it is expected that such genes might be co-expressed in the datasets at hand. 
We have seen that Lasso procedures that are based on the $L1$-norm tend to shrink the number of significant predictors of the model. Unfortunately, such procedures also tend to select only one or a few in a group of highly correlated predictors. Approaches that rely on the $L2$-norm do not entirely solve the issue, since they select all or none of them, increasing false positive or false negatives, respectively. 
In \cite{hebiri2012correlations} it is shown that correlation within the data can consistently influence the Lasso prediction. An important finding regards the relation between correlation and the shrinkage factor $\lambda$ of Equation \ref{eq:lasso}: high correlations tend to lead to smaller tuning parameters. The ability to optimally estimate $\lambda$ by cross-validation does not hold anymore with the presence of highly correlated variables.

\paragraph{Deviation from normality}
The use of penalised regression to infer graphical models of associations between predictor variables, has become increasingly popular after the work published in \cite{glasso, Meinshausen06highdimensional, finegold}. The core idea of such methods is to provide a solution to the variable selection problem by inferring a graph of conditionally dependent predictors. When the complexity of data is also determined by a phenomenon that statisticians call \emph{deviation from normality}, inference and predictions can be significantly impacted by it.
Specifically to the problem of inferring a graph of interactions, contamination of a number of variables can lead to a drastically wrong graph \cite{finegold}.

\paragraph{Degrees of freedom}
As previously mentioned in Section \ref{methods}, the main purpose of penalised regression methods is to reduce the variance of the estimated predictors while controlling the bias by minimising the training error. The best performance is usually reached when an optimal compromise between error and degrees of freedom is found. 
One should expect an increase of the error while decreasing the number of predictors. However, there are counter examples in which more regularisation can, in fact, increase the degrees of freedom. In such cases the regularisation can raise both the error and the degrees of freedom. Examples for Lasso and ridge regression are provided in \cite{kaufman}. 

\subsection{Ridge Regression}
Changing the $L1$-norm to the $L2$-norm in the penalty term of Equation \ref{eq:lasso} is referred to as ridge regression or basis pursuit. The convex optimisation problem to be solved is 

\begin{equation}
\label{eq:ridge}
    \hat{\Theta}^{a,\lambda} = 
    \argmin_{%
      \substack{%
        \text{s.\,t.}\, \Theta:\Theta_a = 0 \\
        \phantom{}\, 
      }
    }
    (\frac{1}{n} \| Y_i - X\Theta \|^{2}_2 + \lambda \| \Theta \|_2)
  \end{equation}

The $L2$-norm has the property of shrinking the regression coefficients without performing selection. Therefore, the number of predictors initially included in the model will stay constant after regression.
Ridge regression regularisation is performed to control the variance of predictors, preventing their coefficients to grow indefinitely. The original motivation for the ridge penalty is to make the problem of regression computable. As a matter of fact, the $\lambda$ shrinkage factor can make the matrix $X^TX$ not invertible, making the calculation of $\beta_{\lambda} = (X^TX + \lambda I_p)^{-1}X^Ty$ not possible \cite{HoerL1}.

The ridge procedure is slightly easier to implement and faster to compute than Lasso.
Generally speaking, ridge regression is to be preferred whenever a high number of minor effects is a realistic hypothesis (even more so, if supported by expert knowledge). 
In contrast, datasets with a small number of significant predictors (main effects) should be regressed with Lasso ($L1$-norm penalty). Whenever this information is available, the choice of the best prediction method is, therefore, straightforward. 

As explained in Section \ref{limitation} a phenomenon that is commonly observed in computational biology is \emph{multicollinearity} \cite{multicollinearity_kvs, farrar1964multicollinearity}, which leads to high variance of the estimator. Ridge regression deals well with highly correlated predictors due to the fact that it will select all of them or none. As a consequence, the mean squared error (MSE) is usually lower than the one of a Lasso procedure. This comes at the cost of including more predictor variables and consequently making the model more complex. 
Ridge regression is best indicated for those applications in which smaller variance is preferred, paying the cost of a more biased prediction. In contrast, in all those procedures that rely on permutation tests to improve the stability of the selected predictors, the ridge penalty is not the best choice, as we will explain in Section \ref{stability}.

Some of the advantages of ridge regression used to discover genetic interactions on simulated and real datasets are illustrated in \cite{Park01012008}. The authors modified the hierarchy rule to add new predictors to the model, allowing an interaction term even in the case in which one of the two genes is present with a strong individual effect. Of course, more complicated rules can be applied, such as those described later in the section that presents hierarchical lasso. Within the same work, a comparison with other tools specifically designed with dimensionality reduction in mind is provided, showing that $L2$-norm penalties usually have reasonable predictive accuracy.

\subsection{Elastic Net}
A method that takes the benefits of both Lasso and Ridge penalties is referred to as \emph{Elastic Net}. The optimisation problem to be solved in this case is 

\begin{equation}
\label{eq:elasticnet}
    \hat{\Theta}^{a,\lambda} = 
    \argmin_{%
      \substack{%
        \text{s.\,t.}\, \Theta:\Theta_a = 0 \\
        \phantom{}\, 
      }
    }
    (\frac{1}{n} \| Y_i - X\Theta \|^{2}_2 + \lambda_1 \| \Theta \|_2 + \lambda_2 \| \Theta \|^{2})
  \end{equation}
  
Elastic net is a method that enforces sparsity, due to the $L1$-norm while favouring the grouping effect of highly correlated predictors, due to the $L2$-norm factor.
The double shrinkage is more demanding in computational terms and more challenging to perform with respect to a pure Lasso or Ridge regression procedure.
%FIXME add examples that use elastic net

\subsection{Fused Lasso}
A generalisation of Lasso that has been designed specifically for predictor variables that can be ordered is referred to as \emph{fused Lasso} \cite{Tibshirani05sparsityand}. 
The core idea of \emph{fused Lasso} consists in penalising the coefficients of the single predictors, as in a regular Lasso procedure, while favouring the sparsity of their differences. 

The optimisation problem to be solved is

\begin{equation}
\label{eq:fusedlasso}
    \hat{\Theta}^{a,\lambda} = 
    \argmin_{%
      \substack{%
        \text{s.\,t.}\, \Theta:\Theta_a = 0 \\
        \phantom{}\, 
      }
    }
    (\frac{1}{n} \| Y_i - X\Theta \|^{2}_2 + \lambda_{1} \sum^{p}_{j=1} \| \theta \|_1 + \lambda_{2} \sum^{p}_{j=2} \| \theta_j - \theta_{j-1} \|)
  \end{equation}
  
\emph{Fused Lasso} is particularly useful for cases in which the number of predictors is much larger than the number of observations ($p >> n$ problems). By penalising the differences of adjacent predictors it is assumed that a limited number of covariates needs to be considered.
One limitation of this approach is that the order of the predictors should be set prior to the regression. Often this information is not available. However, an estimate can be computed directly from the data, i.e. by ordering genes via a correlation metric and applying hierarchical clustering to mark predictors of the same group as adjacent nodes of a graph. This strategy can be also applied to the Group Lasso procedure described in the next section.
The presence of the double penalty factors $\lambda_1$ and $\lambda_2$ requires a more demanding cross-validation procedure, in order to optimally estimate both the parameters.
Researchers have applied \emph{fused Lasso} to synthetic and real data sets with number of predictors in the range between $10^2$ to $4\times10^4$. A direct comparison with $L1$-norm Lasso shows that \emph{fused Lasso} slightly contains the test error while improving overall  sensitivity (true positive rate) \cite{Tibshirani05sparsityand}. In the same simulation study of  controlled predictors, \emph{fused Lasso} does not seem to improve the specificity (true negative rate), compared to the original Lasso procedure. 
Results from a real dataset of Leukemia microarray of $7129$ genes show how fused Lasso reduces the degrees of freedom of the model to $37$ significant genes and performs with the smallest test error, compared to other methods. However, an observation is needed to better frame the consistent improvements of the \emph{fused Lasso} solution. The initial $7125$ genes have been filtered down by a variance-based measure to $1000$. Hierarchical clustering applied to the filtered set of genes has been used to estimate the initial order. As a matter of fact, the pure fusion estimate, without any filtering, performs at the worst, as authors show in the same work.
Another limitation of \emph{fused Lasso} is computational speed, which becomes less practicable for a number of predictors higher than $2000$.

%FIXME add at least two more works of fused lasso in genetics

\subsection{Group Lasso}
Group Lasso, proposed in \cite{Yuan06modelselection} considers the $n-$dimensional vector of responses $Y$ and $p$ predictors which are divided into $L$ groups.
This version of Lasso solves the convex optimisation problem 

\begin{equation}
\label{eq:glasso}
    \hat{\Theta}^{a,\lambda} = 
    \argmin_{%
      \substack{%
        \text{s.\,t.}\, \Theta:\Theta_a = 0 \\
        \phantom{}\, 
      }
    }
    (\frac{1}{2} \| Y_i - \sum^{L}_{l=1} X_l \Theta_l \|^{2}_2 + \lambda \sum^{L}_{l=1} \sqrt n_l \| \Theta_l \|_2)
  \end{equation}

where $n_l$ is the number of predictors per group and the $\sqrt n_l$ takes into account the group size. Group Lasso performs selection at the group level, namely an entire group of genes will be selected or discarded. It comes without saying that a critical aspect of the Group Lasso consists in selecting the groups beforehand. This complicates the tuning even further compared to a regular Lasso in which only parameter $\lambda$ needs to be estimated. Moreover, there is no cross-validation procedure to learn an optimal set of groups, making Group Lasso more challenging when this information is not available.
Another feature worth mention is that the Group lasso procedure does not provide sparsity within the group due to the $L2-$norm in the penalty. However, sparsity can be re-established by another version of the penalty such as 

\begin{equation}
\label{eq:sparseglasso}
    \hat{\Theta}^{a,\lambda} = 
    \argmin_{%
      \substack{%
        \text{s.\,t.}\, \Theta:\Theta_a = 0 \\
        \phantom{}\, 
      }
    }
    (\frac{1}{2} \| Y_i - \sum^{L}_{l=1} X_l \Theta_l \|^{2}_2 + \lambda_1 \sum^{L}_{l=1} \sqrt n_l \| \Theta_l \|_2) + \lambda_2 \| \Theta \|_1
  \end{equation}

in which the $L1-$norm will shrink to zero predictors of the same group. The choice of the sparsity factor requires to optimally estimate an additional parameter $\lambda_2$, usually performed with cross-validation.

A relevant application of the Group Lasso approach has been performed by Friedman et al. in \cite{ggraph}. Starting from the work described in \cite{Meinshausen06highdimensional}, they propose a \emph{symmetrised} version of it, which they call \emph{symmetric lasso}. 
In addition, they adapt the original version of the grouped lasso described in \cite{Yuan06modelselection}, in order to estimate sparse graphical models. The penalty proposed by Friedman et al. groups all of the edges connected to a given node, obtaining a graph that is sparse in its nodes, not in its edges. 
The convex optimisation problem that they try to solve is like the one in Equation \ref{eq:glasso} with groups of equal sizes. The minimisation of the loss and penalty function is performed by means of block-wise coordinate descent.

Another method based on Equation \ref{eq:glasso} is called \emph{paired grouped lasso}, given by the minimisation of  

\begin{equation}
\label{eq:pairedglasso}
    \hat{\Theta}^{a,\lambda} = 
    \argmin_{%
      \substack{%
        \text{s.\,t.}\, \Theta:\Theta_a = 0 \\
        \phantom{}\, 
      }
    }
    (\frac{1}{2} \| Y_i - \sum^{L}_{l=1} X_l \Theta_l \|^{2}_2 + \lambda_1 \sum_{j < i} \| \Theta_{ij}, \Theta_{ji} \|_2)
  \end{equation}
  
with the diagonal elements $\Theta_{ii} = 0$. The overall performance of paired grouped lasso is reported as the best with respect to the original version of grouped lasso as well as the symmetric version. 

\subsection{Hierarchical Lasso}
The problem of detecting pairwise interactions between predictors has received a lot of attention in recent years. When the number of predictors is large, the number of potential interactions grows exponentially with the order of the interaction itself. As a recall, the number of $k$-order interactions from $p$ predictors is $p \choose k$. 

A strategy used by researchers in order to mitigate the curse of dimensionality, consists in testing the interactions of those predictors that show significant individual effects, discarding those that do not. It turns out that choosing the correct threshold for main effects is a not-well-defined problem.

Jacob Bien et al. \cite{bie13las} provide a convex formulation that models main effects and interactions together in hierarchical fashion. The method is an extension of Lasso that incorporates pairwise interactions in order to explain the cases in which 1) two or more genes are expressed together and 2) their contribution to explaining the response in not simply additive.
Starting from a non-hierarchical approach, that they call \emph{all-pairs lasso}, and that we report in Equation \ref{eq:nhlasso}

\begin{equation}
\label{eq:nhlasso}
    \hat{\Theta}^{a,\lambda} = 
    \argmin_{%
      \substack{%
        \text{s.\,t.}\, \Theta:\Theta_a = 0 \\
        \phantom{}\, 
      }
    }
    (\frac{1}{2} \| \sum^{n}_{i=1} Y_i -  X^T_i \beta - \frac{1}{2} X^T_i \Theta X_i \|^{2} + \lambda_1 \| \beta \|_1 + \frac{\lambda}{2} \| \Theta \|_1)
  \end{equation}
  
they extend it by splitting the main effects $\beta_j$ as $\beta^{+}_j - \beta^{-}_j$ 
and by an additional constraint  $ \| \Theta_j \|_1 \leq \beta^{+}_j + \beta^{-}_j$.

The former transformation replaces a non-convex version with a convex relaxation. The latter constraint emphasises the hierarchy of the interactions: the regression coefficient of the $j$-th interaction is a lower bound of the main effects on predictor $j$. It is shown that hierarchy favours models that tend to reuse measured variables. 
In a direct comparison between \emph{hierarchical lasso} and \emph{all-pairs lasso}, it is shown that \emph{parameter sparsity}, defined as the number of nonzero parameters in the model, does not change between the two approaches. On the other hand, \emph{practical sparsity}, defined as the smallest number of variables needed to make predictions, is always lower in hierarchical lasso. An illustrative example is shown in \cite{bie13las}. 

As a conclusion, the degrees of freedom of the hierarchical model is always lower than the ones introduced by Lasso. 
Regardless the useful simplification of the model, hierarchical lasso becomes prohibitive with a relatively high number of predictors. For instance, a network of $1000$ genes, SNPs or proteins will give rise to $4995004$ potential interactions. Considering that a network with $10^6$ nodes is considered only relatively large in biology, it comes without saying that exploring all possible interactions is not feasible with such an approach.
 %\subsection{Nonlinear Lasso}
%Work by Yamada \cite{YamadaJSXS14}

\subsection{Stability of gene regulatory networks}\label{stability}
Inferring genetic interactions within a high dimensional context is a challenging task that often has to deal with the problem of sensitivity. As the number of predictors is increased the true positive rate of all Lasso-based methods explained so far decreases until it approaches a predictive performance comparable to random guessing. Several methods have been designed to deal with the issue of high dimensionality.

The work described in \cite{2012arXiv1206.6519S} performs a permutation-based procedure to test marginal interactions with Lasso regression of a binary response. The significance of each interaction is compared against a null distribution built with $A$ permutations of the response variable $y$ and re-calculating a new set of $\frac{p(p-1)}{2}$ statistics. The expected number of false rejections is computed by taking the average number of these statistics that lie above a cutoff value. The authors call this procedure \emph{TMIcor}, which stands for Testing Marginal Interactions correlation. When applied to real data of Colitis gene expression profiles, the initial $22283$ genes are filtered and only the genes on chromosomes 5 and 10 are considered, as they are known to be related to Crohn's disease. In total, only $663$ genes and $219453$ pairwise interactions are considered. Despite the consistent reduction of the overall dimensionality, the authors claim that \emph{TMIcor} better controls False Discovery Rate, compared to logistic regression.

Another method that leverages the strengths of penalised regression for sparse network structures is described in \cite{robustmethods}. Inspired by the node wise lasso approach of \cite{Meinshausen06highdimensional}, they use a Huber or Least Absolute Deviation (LAD) loss function together with an $L1$-norm penalty to encourage sparsity. A number of approaches, such as tLasso, GLasso, Adalasso, AdaLAD, AdaHuber and Copula, are compared in a simulation study with a number of predictors up to $p=100$. However, a performance evaluation on real data is facilitated by the fact that the authors restrict their attention to $8$ genes already known to be associated to the regulation under study. Another result is collected by performing the list of Lasso-based methods on a dataset with different degrees of contamination. Specifically, the data are generated by a $N(0,\Theta)$ distribution and the contamination occurs by using a $N(\mu, 0.2)$ distribution for different numbers of predictors. Interestingly, the Copula method shows stable performance across various degrees of contamination.

The authors of LABNet \cite{labnet} present a Lasso-based approach to detect main genetic interactions from gene expression profiles. They leverage penalised regression together with a permutation-based procedure that determines whether the predicted interactions are stable across experiments. It is shown that the higher number of permutations not only improves the sensitivity of the method by reducing the number of false negatives, but it also affects the overall number of predicted edges. Unfortunately, a high number of permutations has high computational burden, making the method prohibitive for genome wide studies.
 
A more robust graphical model of gene networks is achieved by using classical and alternative $T-$distributions in \cite{finegold}. In their work, the authors demonstrate that penalised likelihood inference combined with an application of the EM algorithm provides a computationally efficient approach to model selection in the t-distribution case.

Regardless of the impossibility for ridge regression procedures to perform variable selection, they have been considered to estimate a regression coefficient for each predictor variable. The purpose of the authors of \cite{21929786} consists in obtaining an estimate of the significance of each ridge regression coefficient. Specifically they develop and evaluate a test of significance for ridge regression coefficients. Using simulation studies, they demonstrate that the performance of the test is comparable to that of a permutation test, with the advantage of reduced computational cost. The p-value trace is plot for several values of the shrinkage parameter, providing an immediate evaluation of both the estimated $\lambda$ and the significance of the predictors.

%% file: comparison.tex
\section{A comparison of penalised regression methods on known gene-interaction networks} \label{comparison}
In order to evaluate benefits and drawbacks of the methods reported thus far, we sampled synthetic microarrays of 15, 50, 100 and 200 genes each from subnetworks created from a template of E.Coli bacterium. Therefore, we applied the nine penalised regression methods described in Section \ref{methods} to the same datasets and inferred the networks of genetic interactions. In order to infer such networks we followed the strategy used in \cite{Meinshausen06highdimensional}, being so far one of the most intuitive approaches. We subsequently compared each inferred networks to the true network at our disposal. Gene Net Weaver \cite{gnw} has been used to generate both the gold standard and datasets. By doing so, the comparison of inferred and true networks becomes straightforward.

\begin{table*}[htdp]
\centering
\caption{Performance of all penalised-regression based methods inferring a network of 15 genes. True indicates the number of edges in the gold standard network; Pred are the edges of the predicted network; TP, FP, TN and FN stand for true positives, false positive, true negatives and false negatives, respectively; MCC is the Matthew Correlation Coefficient; TPR, FPR and ACC are the true positive rate, false positive rate and accuracy; Time is the amount of seconds to perform the computation.}
\begin{tabular}{|p{2cm}|l|l|l|l|l|l|l|l|l|l|l|} 
\hline
\textbf{}           & \textbf{True} & \textbf{Pred} & \textbf{TP} & \textbf{FP} & \textbf{TN} & \textbf{FN} & \textbf{MCC} & \textbf{TPR} & \textbf{FPR} & \textbf{ACC} & \textbf{Time {[}sec{]}} \\ \hline
\textbf{fused}      & 13            & 24            & 0           & 24          & 188         & 13          & 0            & 0            & 0.11         & 0.83         & 1.17                    \\ \hline
\textbf{hier}       & 13            & 24            & 3           & 21          & 191         & 10          & 0.099        & 0.23         & 0.09         & 0.86         & 7.55                    \\ \hline
\textbf{group}      & 13            & 24            & 1           & 23          & 189         & 12          & 0            & 0.08         & 0.10         & 0.84         & 1.51                    \\ \hline
\textbf{LABnet} & 13            & 24            & 4           & 20          & 192         & 9           & 0.16         & 0.31         & 0.09         & 0.87         & 12.6                    \\ \hline
\textbf{ridge perm} & 13            & 0             & 0           & 0           & 212         & 13          & NA           & 0            & 0            & 0.94         & 12.20                   \\ \hline
\textbf{enet perm}  & 13            & 24            & 2           & 22          & 190         & 11          & 0.037        & 0.15         & 0.10         & 0.85         & 12.87                   \\ \hline
\textbf{lasso}      & 13            & 24            & 0           & 24          & 188         & 13          & 0            & 0            & 0.11         & 0.83         & 0.17                    \\ \hline
\textbf{ridge}      & 13            & 24            & 0           & 24          & 188         & 13          & 0            & 0            & 0.11         & 0.83         & 0.16                    \\ \hline
\textbf{enet}       & 13            & 24            & 0           & 24          & 188         & 13          & 0            & 0            & 0.11         & 0.83         & 0.14                    \\ \hline
\end{tabular}
\label{table15}
\end{table*}

\begin{table*}[htdp]\label{table50}
\centering
\caption{Performance of all penalised-regression based methods inferring a network of 50 genes. Same acronyms as in Table \ref{table15} }
\begin{tabular}{|l|l|l|l|l|l|l|l|l|l|l|l|}
\hline
                    & \textbf{True} & \textbf{Pred} & \textbf{TP} & \textbf{FP} & \textbf{TN} & \textbf{FN} & \textbf{MCC}        & \textbf{TPR}      & \textbf{FPR}       & \textbf{ACC} & \textbf{Time{[}sec{]}} \\ \hline
\textbf{fused}      & 48            & 252           & 5           & 247         & 2205        & 43          & 0.001 & 0.10 & 0.100  & 0.884        & 177.66       \\ \hline
\textbf{hier}       & 48            & 224           & 21          & 203         & 2249        & 27          & 0.170   & 0.4375            & 0.08   & 0.908        & 544.71       \\ \hline
\textbf{group}      & 48            & 414           & 21          & 393         & 2059        & 27          & 0.102   & 0.4375            & 0.16  & 0.832        & 17.84       \\ \hline
\textbf{LABnet} & 48            & 98            & 16          & 82          & 2370        & 32          & 0.212   & 0.33 & 0.03  & 0.9544       & 206.2       \\ \hline
\textbf{ridge perm} & 48            & 0             & 0           & 0           & 2452        & 48          & NA                  & 0                 & 0                  & 0.9808       & 202.60       \\ \hline
\textbf{enet perm}  & 48            & 86            & 10          & 76          & 2376        & 38          & 0.133   & 0.20 & 0.030 & 0.9544       & 201.93       \\ \hline
\textbf{lasso}      & 48            & 254           & 8           & 246         & 2206        & 40          & 0.030  & 0.16 & 0.10  & 0.8856       & 1.345       \\ \hline
\textbf{ridge}      & 48            & 254           & 8           & 246         & 2206        & 40          & 0.030  & 0.16 & 0.10  & 0.8856       & 1.28        \\ \hline
\textbf{oneenet}    & 48            & 254           & 8           & 246         & 2206        & 40          & 0.030  & 0.16 & 0.10  & 0.8856       & 1.52       \\ \hline
\end{tabular}
\end{table*}

\begin{table*}[htdp]
\centering
\caption{Performance of all penalised-regression based methods inferring a network of 200 genes. Same acronyms as in Table \ref{table15}}
\begin{tabular}{|l|l|l|l|l|l|l|l|l|l|l|l|}
\hline
                     & \textbf{True} & \textbf{Pred} & \textbf{TP} & \textbf{FP} & \textbf{TN} & \textbf{FN} & \textbf{MCC}         & \textbf{TPR}       & \textbf{FPR}       & \textbf{ACC} & \textbf{Time{[}sec{]}} \\ \hline
\textbf{fused}      & 212       & 398     & 76           & 322         & 39466      & 136    & 0.256  & 0.358 & 0.008 & 0.98    & 10761   \\ \hline
\textbf{hier}       & 212        & 428    & 95    & 333         & 39455   & 117       & 0.310    & 0.448  & 0.008  & 0.988   & 72000  \\ \hline
\textbf{group}      & 212      & 432    & 3             & 429         & 9670        & 78       & 0.046   & 0.103  & 0.024 & 0.96  & 734        \\ \hline
\textbf{LABnet} & 212   & 398          & 103      & 295         & 39493        & 109   & 0.349    & 0.485  & 0.007 & 0.98  & 3650   \\ \hline
\textbf{lasso}      & 212      & 400          & 38            & 362        & 39426     & 174    & 0.124 & 0.179 & 0.009 & 0.986   & 43.5    \\ \hline
\textbf{ridge}      & 212      & 400           & 38            & 362        & 39426     & 174    & 0.124 & 0.179 & 0.009 & 0.986   & 48.2    \\ \hline
\textbf{enet}       & 212     & 400           & 38           & 362         & 39426    & 174      & 0.124 & 0.179 & 0.009 & 0.986  & 50.4     \\ \hline
\end{tabular}\label{table200}
\end{table*}

All the methods described thus far have been applied directly to the datasets at hand, with the exception of Group Lasso and Fused Lasso. As previously mentioned these two methods require the predictor variables to be ordered according to some criterium in order to apply the convex optimisation described by Equation \ref{eq:glasso} and Equation \ref{eq:fusedlasso}. 
In both cases, we group all predictors by a correlation metric. Specifically, for each regression iteration, we build the correlation matrix $C$ of the $p-1$ predictors. We then perform hierarchical clustering on $C$, and generate $k$ clusters by using the Euclidean distance measure. The optimal number of clusters has been empirically estimated to $k=3$ for Group Lasso and $k=10$ for Fused Lasso.
Building the vector of grouped predictors directly from the output of the hierarchical clustering procedure is straightforward. Finally, we use a 10-fold cross-validation to estimate the penalty factor $\lambda$. We are aware of the fact that a better grouping metric might exist. However, without any prior knowledge, this metric is challenging to obtain or infer. Performance benchmarks with timing measures are reported in Table \ref{table15}, Table \ref{table50} and Table \ref{table200}. 

The number of predicted edges has been normalised across all methods with a quantile-based selection that filters out small regression coefficients. This normalisation procedure allows all the methods to predict a comparable number of edges. 

The method that performs the best in the 15-gene network is \emph{LABnet} \cite{labnet}, a mixture of $L1-$norm Lasso and permutation test that increases the stability of the inferred topology. The number of predicted edges of \emph{LABnet} is also one of the lowest, improving its overall performance, measured by the Matthew Correlation Coefficient. As expected and already mentioned by its authors, the permutation test of \emph{LABnet} is heavily detrimental with ridge regression and slightly less with elastic net due to the fact that $L2$-norm penalties do not shrink to zero the regression coefficients. 

\emph{LABnet} performs with the highest MCC in the 50-gene dataset too. Hierarchical Lasso performs equally like with a computational overhead 2x as larger. This is due to the fact that hierarchical Lasso performs regression on $p^2$ predictor variables (it considers all pairwise interactions).

\emph{LABnet} is also the best method in the 200-gene network with $MCC=0.349$. Hierarchical Lasso has comparable performance with execution time 20x as higher than \emph{LABnet}. As expected the computational burden of hierarchical Lasso is exponential in the number of predictors. Regardless the undeniable computational burden required by hierarchical Lasso to regress a quadratic number of covariates, \emph{hierNet} - the R package that implements it - has not been designed with parallelisation in mind. On the other side, \emph{LABnet} takes advantage of multi core processor to parallelise the permutation-based procedure. 

%% file: discussion.tex
\section{Discussion}
Despite active research in the field of genetic interaction networks by penalised regression methods, many limitations still need to be addressed. More sophisticated solutions need to be provided when dealing with high dimensional data and highly correlated variables. Multicollinearity is a recurrent problem in genetics and, according to the relatively poor performance of the methods reported thus far, simply applying penalised regression does not seem to provide acceptable solutions. 
All the methods reported in this review have been applied to genetic datasets with a limited number of covariates. Our primary goal is to provide an unbiased comparison among all the methods under investigation. We are aware that preselecting variables might improve the stability and the overall performance in terms of prediction and reduce computation time. Preselection, however, is an open problem in genetics and it can lead to complete removal of detectable signals or biased results, depending on the strength of the preselection filter.
Even in the case of moderate correlation, quite rare in biology, applying penalised regression within a high dimensional context might become prohibitive. All permutation-based procedures impose a computational burden that makes them impossible to be considered for real world datasets (number of predictors approaching $10^9$). 
%In such cases, a parallel computing infrastructure is essential in order to perform genome wide association studies. 
% this is your suggestion based on other work not good to mention here We address data integration of heterogeneous sources as a valuable mitigation strategy to the problem of high dimensionality. 
The limitations of penalised regression methods suggest that constraining the problem of variable selection with diverse datasets or with prior knowledge can reduce the variance of the predictions and, in turn, increase their significance.

%% file: conclusion.tex
\section{Conclusion}\label{conclusion}
We provided a review of nine penalised regression methods applied to gene expression data to infer the topology of the network of gene-gene interactions. The different types of penalties are indicated within diverse contexts, according to initial hypotheses of strong presence of main effects rather than weak interactions.
We found a limitation that is common to all approaches and that regards high dimensionality and multicollinearity within the datasets at hand. None of the methods described seem to deal with both at the same time. Due to the nature of genetic data - high number of highly correlated variables - we suggest to consider penalised regression only in more constrained problems, where prior knowledge and data integration play a fundamental role. As a consequence, more sophisticated regression-based approaches need to be designed to make the prediction of gene regulatory networks more reliable.